\documentclass[preprint]{ptptex}

\newcommand{\bra}{\langle}
\newcommand{\ket}{\rangle}

\newcommand{\del}{\partial}

\notypesetlogo  
\preprintnumber[3cm]{KYUSHU-HET-66}

\title{
Neutrino Masses and Mixing Matrix from \mathversion{bold}$SU(1,1)$ Horizontal Symmetry
}

\author{
Kenzo INOUE \footnote{E-mail: inou1scp@mbox.nc.kyushu-u.ac.jp}
and Nao-aki YAMASHITA \footnote{E-mail: naoaki@higgs.phys.kyushu-u.ac.jp}
}

\inst{
Physics Department, Kyushu University,\\ Fukuoka 812-8581, Japan
}

\abst{
A new mechanism is proposed to explain neutrino masses and their mixing via $SU(1,1)$ horizontal 
symmetry breaking. Based on this mechanism, the smallness of neutrino masses is realized and 
a hierarchical spectrum with large mixing is given. 
}

\begin{document}
\maketitle
Progress in several neutrino experiments has gradually revealed 
the generational structure of neutrino.\cite{review1,review2}
In addition to the smallness of their masses,
the mixing matrix, the so-called Maki-Nakagawa-Sakata (MNS) matrix, \cite{MNS} is 
established as a bi-large mixing, in contrast to the Cabbibo-Kobayashi-Maskawa (CKM) matrix.\cite{CKM}  
It is important that neutrinos have a generational 
structure that is quite different from the structure of quarks   
given by the hierarchical Yukawa couplings.
There seems to be no satisfactory phenomenological model 
which can give a comprehensive understanding of these two different generational structures.

Horizontal symmetry, as a symmetry between the generations,  
is expected to give a unified view of
 the generational structure of quarks and leptons.
One of the authors has proposed  
a model based on the noncompact horizontal symmetry group $G_{\rm H}=SU(1,1)$,\cite{KI1,KI2}
which gives the characteristic hierarchical Yukawa couplings.
The model is a vector-like extension of the minimal supersymmetric standard model (MSSM),
in which a horizontal  multiplet is an infinite dimensional unitary representation of $SU(1,1)$.
All MSSM fields that have the same gauge quantum numbers are embedded into a horizontal multiplet
$F$.
A partner $\bar{F}$  is additionally introduced to make the theory vector-like for each horizontal
multiplet $F$.
From infinite components of horizontal multiplets $\{F,\bar{F}\}$,  
a finite number of chiral generations are generated through
the spontaneous breakdown of horizontal symmetry.
This mechanism is called the `spontaneous generation of generations'.
It causes all trilinear couplings of all chiral generations to be hierarchical. 
Thus hierarchal Yukawa couplings appear naturally.
Based on this mechanism, reasonable mass ratios of quarks and charged leptons and 
the CKM matrix have been explained.

In addition, the model provides various phase structures through the horizontal symmetry breaking.\cite{KY}
Various phases are characterized by  order-one parameters $r_{F}$ and $r_{g}^{\rm c}$, 
where $r_{F}$  is a quantity related to 
 $F$-$\bar{F}$ couplings and a symmetry breaking pattern, and $r_{g}^{\rm c}$  is a critical value.
In a phase that satisfies the condition $r_{F} < r_{g}^{\rm c}$, $g$ chiral generations appear from 
horizontal multiplets $\{F,\bar{F} \}$.
In a phase satisfying $r_{F} > r_{g}^{\rm c}$, there are no chiral generations and all become massive.
According to the decoupling of massive components, various types of chiral generational structure 
appear at a low energy scale.  
The advantage of the spontaneous generation of generations is 
that the phase structure is controlled by order-one parameters.
This property is highly compatible with  grand unified theory (GUT), 
because superfluous GUT partners, $\{ F',\bar{F}'\}$, can become super heavy without fine turning. 
To realize this splitting between $F$ and $F'$ is to obtain a natural setup 
in which the conditions  $ r_{F'} > r_{g}^{\rm c} > r_{F}$ are satisfied. 
The phase structure is determined by order-one parameters, 
and this implies that $SU(1,1)$ horizontal symmetry would have an important role in GUT.

 When considering the neutrino masses and mixings based on  $SU(1,1)$ horizontal symmetry,
it is not possible to realize an extension based on the see-saw mechanism\cite{seesaw} in a natural way,
 because $SU(1,1)$ symmetry allows a Dirac mass $F{\bar F}$ but  forbids a Majorana mass term $FF$ 
in the superpotential.
Therefore we need a different extension in the horizontal symmetry approach
to give desirable structure to neutrinos. 
In this paper we try to give a reasonable structure of masses and mixings for neutrinos 
based on the formalism described in Ref \ref{KY}). 
As a simple extension, we introduce additional horizontal multiplets $\{T_{-\beta},\bar{T}_{\beta} \}$ represented by
\begin{eqnarray}
T_{-\beta} = \{ t_{-\beta}, t_{-\beta-1}, t_{-\beta-2},  \cdots \},\qquad
\bar{T}_{\beta} = \{ \bar{t}_{\beta}, \bar{t}_{\beta+1}, \bar{t}_{\beta+2},  \cdots \},
\end{eqnarray}
where $-\beta$ $(<0)$ and  $\beta$ are the highest and lowest weights of
the infinite dimensional unitary representations of $SU(1,1)$.
We  assign a representation $(\mathbf{1},\mathbf{3}  ,\mathbf{+1} )$ of the standard model (SM)  gauge group 
$SU(3)\times SU(2)\times  U(1)$ to $T_{-\beta}$.\cite{triplet} 
Following the vector-like approach, $\bar{T}_{\beta}$ is assigned to the conjugate representation of $T_{-\beta}$.
We consider the generation generating superpotential 
\begin{equation}
u_{T}\Phi_R T_{-\beta} \bar{T}_{\beta} +v_{T}\Psi_R T_{-\beta} \bar{T}_{\beta}, \label{eq:Tmass}
\end{equation}
where $\Phi_R =\{ \phi_{i} \}_{i=-R}^{R} $ and $\Psi_R=\{ \psi_{i} \}_{i=-R}^{R}$ 
are SM gauge singlets, and  the non-unitary finite dimensional representations of the horizontal group,
$u_{T}$ and $v_{T}$, are complex coupling constants.
We assume that both $\Phi_R $ and $ \Psi_R $ acquire non-vanishing vacuum expectation values (VEVs) 
$\bra \phi_{0} \ket$ and $\bra \psi_{g} \ket$, which break the horizontal symmetry.

For quarks and leptons, 
considering the extension to the unification of the GUT group and the horizontal symmetry group $G_{\rm H}$, 
we assign different positive $SU(1,1)$ weights, $\alpha_{f= Q,L,U,D,E}$. 
 For the same reason, two Higgs doubles in MSSM are assigned different negative $SU(1,1)$ weights, $-\rho_{f= U,D}$.
We denote the MSSM superfields as $F=\{ F_{+} ,F_{-}\}$, where 
\begin{eqnarray}
F_{+} &=& \{ Q_{\alpha_{Q}},~L_{\alpha_{L}},~\bar{U}_{\alpha_{U}},~\bar{D}_{\alpha_{D}},~\bar{E}_{\alpha_{E}}\},\\
F_{-} &=&\{ H_{-\rho_{U}},~H^{'}_{-\rho_{D}} \}.
\end{eqnarray}
Each MSSM superfield has a vector partner.
The parters are denoted as $\bar{F}=\{ \bar{F}_{+} ,\bar{F}_{-}\}$, where 
$\bar{F}_{+} =\{ \bar{Q}_{-\alpha_{Q}},~\bar{L}_{-\alpha_{L}},~{U}_{-\alpha_{U}},
~{D}_{-\alpha_{D}},~{E}_{-\alpha_{E}}\}$ 
and $\bar{F}_{-} = \{\bar{H}_{\rho_{U}},~\bar{H'}_{\rho_{D}} \}$. 
All components in $\bar{F}$ decouple from the low energy world through horizontal symmetry breaking.
These quantum number assignments allow a trilinear superpotential which consists of unitary fields as follows:
\begin{equation}
\mathcal{W}_{F} + \mathcal{W}_{\bar{F}} + \mathcal{W}_{T} + \mathcal{W}_{\bar{T}},\label{all}
\end{equation}
where 
\begin{eqnarray}
\mathcal{W}_{F} &=& y^{e}H'_{-\rho_{D}}L_{\alpha_{L}}\bar{E}_{\alpha_{E}} 
+ y^{d}H'_{-\rho_{D}}Q_{\alpha_{Q}}\bar{D}_{\alpha_{D}}+ y^{u}H_{-\rho_{U}}Q_{\alpha_{Q}}\bar{U}_{\alpha_{U}
},\label{W_F}
\\
\mathcal{W}_{T} &=& w^{(1)}L_{\alpha_{L}}T_{-\beta}L_{\alpha_{L}} + 
w^{(2)}\bar{H}_{\rho_{U}}T_{-\beta}\bar{H}_{\rho_{U}} + 
w^{(3)}\bar{H}_{\rho_{U}}T_{-\beta}L_{\alpha_{L}}.  \label{W_T}
\end{eqnarray}
Here, $\mathcal{W}_{\bar{F}}$ and  $\mathcal{W}_{\bar{T}}$
are conjugates of $\mathcal{W}_{F}$ and  $\mathcal{W}_{T}$, respectively.
This corresponds to replacing $\{F,\bar{F}, T, \bar{T} \}$ to $\{\bar{F}, F, \bar{T}, T\}$ in each superpotential. 
The coefficient $y^{e,d,u}$ and  $w^{(i=1,2,3)}$ are complex numbers.
Note that there is no $R$ parity broken term in these superpotentials. 
To realize the Yukawa hierarchy in $\mathcal{W}_{F}$, we must assign $SU(1,1)$ weights that  satisfy the 
 relations\cite{KY}
\begin{eqnarray}
&&\alpha_{Q} + \alpha_{U} = \rho_{U},\\
&&\alpha_{Q} + \alpha_{D} = \alpha_{L} +\alpha_{E} = \rho_{D}. \label{eq:hu}
\end{eqnarray}
In addition,  the $SU(1,1)$ invariance requires other conditions which involve $\beta$.
As an example, we consider a trilinear coupling $H\bar{T}H$ in $\mathcal{W}_{\bar{T}}$.   
The coupling $H\bar{T}H$ is described by   
\begin{equation}
H_{-\rho_{U}}\bar{T}_{\beta}H_{-\rho_{U}} = \sum_{i,j=0}^{\infty}B^{H}_{i,j}h_{-\rho_{U} -i} \bar{t}_{\beta +i+j -\Delta'} h_{-\rho_{U} -j},
\end{equation}
where $B^{H}_{i,j}$ is a Clebsh-Gordan (CG) coefficient of  $H\bar{T}H$ 
whose explicit form is 
\begin{eqnarray}
&&B^{H}_{i,j} = N^{H} (-1)^{i+j}
\sqrt{\frac{i!j!\Gamma(4\alpha + i)\Gamma(4\alpha + j)}{(i+j-\Delta')!\Gamma(2\beta + i+j -\Delta')}}\nonumber \\
&&\qquad \times \sum_{r = 0}^{\Delta'} 
\frac{(-1)^{r} (i +j-\Delta')!}{(i-r)!(j+r-\Delta')!r!(\Delta'-r)!\Gamma(4\alpha + r)\Gamma(4\alpha -r +\Delta')}.
\label{CG-B} \\
&& (\mbox{with $N^{H}$ a  normarilzation constant and  $\Delta' =  \beta- 2 \rho_{U}$ a non-negative integer})
\nonumber  
 \end{eqnarray}
The CG coefficient $B^{H}_{i,j}$ has the property $B^{H}_{i,j} = ( -1)^{\Delta'}B^{H}_{j,i}$. 
To realize a non-vanishing coupling, 
$B^{H}_{i,j}$ must be symmetric with respect to $i$ and $j$  in Eq. (\ref{CG-B}).
Thus we assume that $\beta - 2 \rho_{U}$ $(=\Delta')$ is a non-negative even integer.
This argument is also applied to the coupling $LTL$. 
It is thus found that 
so that $\beta - 2 \alpha_{L}$ $(=\Delta)$ must be a non-negative even integer.

Suppose that all components of $T$ and $\bar{T}$ are massive  but that $F$ has appropriate chiral generations
obtained through the spontaneous generation of generations as
\begin{eqnarray}
&&T \longrightarrow T_{{\rm massive}}
,\qquad \bar{T} \longrightarrow \bar{T}_{\rm massive},\label{decomposition1} \\
&&F\longrightarrow  F_{\rm massless}+F_{{\rm massive}}
,\qquad \bar{F} \longrightarrow \bar{F}_{\rm massive}, \label{decomposition2}
\end{eqnarray}
where $F_{\rm massless}$ represents MSSM superfields, 
so that $F_{+}$ has three chiral generations and $F_{-}$ has one chiral generation. 
Each chiral generation is induced by generation generating superpotentials with non-unitary fields  
$\Phi_{R_{\pm}}^{(\pm)} = \{ \phi^{(\pm)}_{i} \}_{i=-R_{\pm}}^{R_{\pm}}$ 
and $\Psi_{R_{\pm}}^{(\pm)}= \{ \psi^{(\pm)}_{i} \}_{i=-R_{\pm}}^{R_{\pm}}$:

\begin{equation}
u_{F_{+}}\Phi^{(+)}_{R_{+}} F_{+} \bar{F}_{+} +v_{F_{+}}\Psi^{(+)}_{R_{+}} F_{+} \bar{F}_{+}, \label{eq:GG+}
\end{equation}
and
\begin{equation}
u_{F_{-}}\Phi^{(-)}_{R_{-}} F_{-} \bar{F}_{-} +v_{F_{-}}\Psi^{(-)}_{R_{-}} F_{-} \bar{F}_{-}.\label{eq:GG-}
\end{equation}
Appropriate chiral generations are realized with the condition
$r_{F_{+}}< r_{3}^{\rm c}$ and $r_{F_{-}}< r_{1}^{\rm c}$, where
\begin{equation}
r_{F_{+}} = \frac{u_{F_{+}}\bra \phi^{(+)}_{0}\ket}{v_{F_{+}}\bra \psi^{(+)}_{-3}\ket} ,
\qquad r_{3}^{\rm c} = \sqrt{\frac{R_{+}(R_{+}-1)(R_{+}-2)}{(R_{+}+1)(R_{+}+2)(R_{+}+3)}}
\end{equation}
and
\begin{equation}
r_{F_{-}} = \frac{u_{F_{-}}\bra \phi^{(-)}_{0}\ket}{v_{F_{-}}\bra \psi^{(-)}_{1}\ket},
\qquad r_{1}^{\rm c} = \sqrt{\frac{R_{-}}{R_{-}+1}}.
\end{equation}
Here, $\bra \phi^{(\pm)}_{0} \ket $, $\bra \psi^{(+)}_{3}\ket $
 and $\bra \psi^{(-)}_{1} \ket$
 are the non-vanishing VEVs of non-unitary fields coupled to $F$ and $\bar{F}$.
The consequence that all components of $T$ and $\bar{T}$ become massive is realized if $r_{T} > r^{\rm c}_{g}$,
where 
\begin{equation}
r_{T} = \frac{u_{T}\bra \phi_{0}\ket}{v_{T}\bra \psi_{g}\ket},
\qquad r_{g}^{\rm c} = \sqrt{\frac{R!R!}{(R-g)!(R+g)!}}.
\end{equation}

Under the splitting between massive and massless components in Eqs. (\ref{decomposition1}) and (\ref{decomposition2}),
the superpotential (\ref{all}) is symbolically described as
\begin{equation} 
\mathcal{W}({\rm massless}) +\mathcal{W}({\rm massless}, {\rm massive}) +\mathcal{W}({\rm massive}). 
\end{equation} 
The low energy dominant term  $\mathcal{W}_{\rm eff}^{(0)}$ is  $\mathcal{W}(\rm massless)$ and   
equal to the standard MSSM Yukawa superpotential ${\mathcal W}_{F}({\rm massless )}$.
In addition, we consider  $\mathcal{W}_{\rm eff}^{(1)}$ at the next order 
It is induced by one massive component exchange, like
\begin{equation} 
\int d\mu_{\rm massive}\mathcal{W}({\rm massless},{\rm massive})\times \mathcal{W}({\rm massless}, {\rm massive}) 
\sim \frac{1}{M} \mathcal{W}_{\rm eff}^{(1)}(\rm massless),
\end{equation} 
where $\int d\mu_{\rm massive}$ represents the procedure of integrating 
out the degree of freedom of massive components, and 
$M$ is the horizontal symmetry breaking scale proportional to the masses of massive components.
At this level, an effective interaction is 
induced by the massive components of $T$ and $\bar{T}$, 
 and the left-handed neutrino mass terms are given by
\begin{eqnarray}
&&\int d\mu_{T\bar{T}}\left(
w^{(1)}{L}_{\rm massless}{T}_{\rm massive}{L}_{\rm massless}\times 
w^{(2)}{H}_{\rm massless}{\bar{T}}_{\rm massive}{H}_{\rm massless}  \right)\nonumber \\
&&\qquad \qquad\qquad \longrightarrow   
\frac{1}{ M_{T\bar{T}}}(  \bra H \ket^{2}L L)_{\rm massless}, 
\label{eq:nu-mass}
\end{eqnarray}
where $M_{T\bar{T}}$ is a typical mass scale of $T$--$\bar{T}$ couplings.
The smallness of neutrino masses is guaranteed by $ \bra H \ket  \ll M_{T\bar{T}}$. 
Note that 
the last term in Eq. (\ref{W_T}) does not contribute,
 and there is no other allowed interaction at this level.

To integrate the massive components of $T$ and $T$,  
we need to specify the $SU(1,1)$ breaking pattern.
For simplicity we consider two cases:
\begin{eqnarray} 
&&\mbox{(i)}~~ (\Phi_{R}, \Psi_{R}) = (\Phi^{(+)}_{R_{+}}, \Psi^{(+)}_{R_{+}}),\label{case1}\\ 
&&\mbox{(ii)}~~ (\Phi_{R}, \Psi_{R}) =(\Phi^{(-)}_{R_{-}}, \Psi^{(-)}_{R_{-}}).\label{case2}
\end{eqnarray} 
The first step of integrating  $T$ and $\bar{T}$ in the supersymmetric way consists of 
solving the following equation for $ t_{-\beta -i}$($\bar{t}_{\beta +i}$): 
\begin{equation}
\frac{\del {\mathcal W'}_{T} }{\del \bar{t}_{\beta+ i}}= 0 ,~~
\left( \frac{\del {\mathcal W'}_{T} }{\del t_{-\beta -i}}= 0 \right)
\label{eq:infeq}
\end{equation}
where
\begin{equation}
{\mathcal W'}_{T} =  u_{T}\bra \Phi_R \ket T_{-\beta} \bar{T}_{\beta} +v_{T}\bra \Psi_R \ket{T}_{-\beta} \bar{T}_{\beta} 
+w^{(1)}L_{\alpha_{L}}T_{-\beta}L_{\alpha_{L}} + 
w^{(2)}{H}_{-\rho_{U}}\bar{T}_{-\beta}{H}_{-\rho_{U}}.  
\label{eq:WforInt}
\end{equation}
The solution  for $ t_{-\beta -i}$($\bar{t}_{\beta +i}$) 
 depends on all components of $H_{-\rho_{U}} ( L_{\alpha_{L}})$. 
However, what we need are terms of massless components, 
because the neutrino mass terms in the low energy effective interaction depend on massless components of $H$ and $L$.  
As a consequence of  the spontaneous generation of chiral generation, 
the decompositions between chiral components ($h$ and $l^{(i=0,1,2)}$)  and the massive components
expressed in Eq. (\ref{decomposition2}) 
 are given by
\begin{eqnarray}
&&h_{-\rho_{U} -i} = a_{i}^{h}h + {\rm massives},\qquad\label{eq:higgs}
l_{\alpha_{L} +i} =\sum_{j= 0,1,2} a_{i}^{l(j)}l^{(j)} + {\rm massives}. \label{eq:lepton}
\end{eqnarray}
The mixing coefficients between the chiral base and the horizontal base are 
\begin{equation}
a_{j}^{h} =r_{H}^{j}a_{0}^{h}\prod_{n=0}^{j-1}\frac{A_{n,n}^{2\alpha,R}}{ A_{n+1,n}^{2\alpha,R}},
\qquad 
a^{l(j)}_{3i+j} =r_{L}^{i}a_{j}^{l(j)}\prod_{n=0}^{i-1}\frac{A_{3n+j,3n+j}^{\alpha,R}}{ A_{3n+j,3(n+1) +j}^{\alpha,R}},
\end{equation}
where $A_{i,j}^{\alpha,R}$ is an  $SU(1,1)$ CG coefficient described in Ref. \ref{KI2}).
The effective part of the solution $t_{-\beta -i}$ of Eq. (\ref{eq:infeq}) is described 
with the Pauli matrix  $\sigma^{a}$  and $\epsilon = i \sigma^{2}$
as 
\begin{equation}
(t^{a}_{-\beta -i})_{\rm massless} = ( h\epsilon \sigma^{a} h){\mathcal T}_{i}.
\end{equation}
The explicit form of $\mathcal T_{i}$ for the case (i) is 
\begin{eqnarray}
&&{\mathcal T_{i}} = \frac{w^{(2)}}{2u_{T}\bra\phi_{0}\ket A^{\beta,R}_{i,i}}\nonumber \\
&&~ \times \sum_{n=i}^{\infty} \left[\frac{1}{(-r_{T})^n}  \left(\prod_{m=1}^{n}  \frac{  A_{i+m-1
,i+m}^{\beta,R}}{A^{\beta,R}_{i+m,i+m}} \right)
\left(\sum_{j=0}^{i+n+\Delta'} B^{H}_{j,i+n-j+\Delta' }a^{h}_{j}a^{h}_{i+n-j+\Delta'}\right)
\right] \label{eq:Ti}
\end{eqnarray}
and for the case (ii) is 
\begin{eqnarray}
&&{\mathcal T}_{3i+k} = \frac{w^{(2)}}{2u_{T}\bra\phi_{0}\ket A^{\beta,R}_{3i+k,3i+k}}
\sum_{n=0}^{i}\left[ \frac{1}{(-r_{T})^{n}}\left( \prod_{m=1}^{n} 
 \frac{ A^{\beta,R}_{3(i-m+1) +k,3(i-m) +k}}{ A^{\beta,R}_{3(i-m) +k,3(i-m)+k}} \right)\right. \nonumber \\
&&\qquad \qquad \left. \times \left(\sum_{j=0}^{3(i-n)+k+\Delta'} 
B^{H}_{j,3(i-n) + k -j +\Delta' }a^{h}_{j}a^{h}_{3(i-n)+ k-j+\Delta'}\right) \right]. (k=0,1,2) \label{eq:Tii}
 \end{eqnarray}

A neutrino mass term appears from massless components of $-w^{(1)}L_{\alpha_{L}}T_{-\beta}L_{\alpha_{L}}$ and 
takes the from
\begin{eqnarray}
&&\qquad\qquad  \left. -w^{(1)}L_{\alpha_{L}}T_{-\beta}L_{\alpha_{L}}\right|_{\rm massless} =
\sum_{r,s = 0,1,2} \Gamma^{\nu}_{r,s}\nu^{(r)}\nu^{(s)}, \label{eq:mass}\\
&&\qquad\Gamma^{\nu}_{r,s} = -w^{(1)}w^{(2)}\bra h \ket^{2} \sum _{i,j = 0}^{\infty} B_{i,j}^{L} {\mathcal T}_{i+j -\Delta} a_{i}^{l(r)}a_{j}^{l(s)}.
\\
&&\qquad\qquad  ~~(\mbox{with $\Delta = \beta -2 \alpha_{L}$ a non-negative even integer})\nonumber
\end{eqnarray}
The CG coefficient $B^{L}_{i,j}$ is obtained by replacing $\Delta'$ 
with $\Delta$ and $\rho_{U}$ with $\alpha_{L}$ in Eq. (\ref{CG-B}).
$\Delta$ and  $\Delta'$  are both non-negative even integers, 
and thus the possible $SU(1,1)$ weight assignments are 
constrained.
From Eq. (\ref{eq:hu}) and the formulas for $\Delta$ and $\Delta'$, 
the $SU(1,1)$ weights $\alpha_{E}$ and $\alpha_{L}$ are given by
\begin{equation}
\alpha_{E} = \rho_{D} -\rho_{U} +(\Delta -\Delta')/2,\qquad
\alpha_{L} = \rho_{U} +(\Delta' -\Delta)/2. \label{eq:weights_relation}
\end{equation}

The neutrino masses  are given by ${\rm diag}(m_{1},m_{2},m_{3}) =   U_{\nu}^{\rm T} \Gamma^{\nu} U_{\nu}$, 
where $U_{\nu}$ is a unitary matrix. 
To investigate the MNS matrix $U_{\rm MNS}=U_{e}^{\dagger}U_{\nu}$ in this framework, 
we must determine  a unitary matrix $U_{e}$ which rotates left-handed charged leptons in a bi-unitary transformation.
We can determine $U_{e}$ so as to give a parameter set that realizes charged lepton mass ratios.
This parameter set also has to realize a reasonable quark mass hierarchy and the CKM matrix simultaneously. 
A quantity in the quark sector is easily reproduced when $\rho_{U}\sim 1$ and $\rho_{D}\sim 1$,\cite{KY}
and therefore we must assign $SU(1,1)$ weights to satisfy these two conditions.
Considering the positivity of $\alpha_{E}$ and $\alpha_{L}$ in Eq. (\ref{eq:weights_relation}),
the condition on $\rho_{D}$ and $\rho_{U}$ requires that $(\Delta -\Delta')/2$ must be $0$ or $1$. 
If we employ the  simple assignment $\rho_{U} = \rho_{D} = \rho$ $(\sim 1)$, 
then $(\Delta -\Delta')/2 =1 $ is allowed. 
While the assumption of minimal $SU(1,1)$ weights, which means assigning the same weight to quarks and leptons,    
and  gives reasonable mass ratios and the CKM matrix, 
the existence of a neutrino mass in this model forbids such a weight assignment.
Thus we must modify the minimal weights assumption while maintaining 
the weights relations Eq. (\ref{eq:hu}) and Eq. (\ref{eq:weights_relation}) 
to realize a reasonable mass hierarchy of quarks and charged leptons. 
With an appropriate setting of the $SU(1,1)$ weights we explore realizations of neutrino mass ratios and the MNS matrix
.

It is difficult to reveal detailed properties of the resulting neutrino mass matrix $\Gamma^{\nu}_{r,s}$
analytically.
For small $\Delta$ and $\Delta'$, however,  it would be naively expected that the entire structure of $\Gamma^{\nu}$
 is controlled by these two parameters. 
This is because the coefficients $a_{i}^{l(r)}$ and $a_{i}^{h}$ in $\Gamma^{\nu}$ 
are monotonically decreasing function of the  index $i$
, which makes higher-mode contributions irrelevant,  and 
the lower-mode contribution  is controlled by the CG coefficient $B_{i,j}$, especially $\Delta$ and $\Delta'$, 
 and is suppressed for large  $\Delta$ and $\Delta'$.
In the case  $\Delta=\Delta'=0$, 
the matrix elements $\Gamma^{\nu}_{r,s}(=\Gamma^{\nu}_{s,r})$ satisfy the hierarchical relations 
$|\Gamma^{\nu}_{2,2}| > |\Gamma^{\nu}_{2,1}| > |\Gamma^{\nu}_{2,0}| \sim |\Gamma^{\nu}_{1,1}| 
> |\Gamma^{\nu}_{1,0}| > |\Gamma^{\nu}_{0,0}|$. This is the same as in the case of the Yukawa matrix.
Because the realized  mass spectrum is hierarchical, like the quark and charged lepton, 
this setting would not  give feasible neutrino mass square differences. 
For $\Delta =2$ and $\Delta'=0$, then we have  
$ |\Gamma^{\nu}_{1,1}| \sim  |\Gamma^{\nu}_{2,0}|  > |\Gamma^{\nu}_{0,1}|> |\Gamma^{\nu}_{0,0}| 
> |\Gamma^{\nu}_{1,2}| >|\Gamma^{\nu}_{2,2}|$. 
In this case, two of the mass eigenvalues are close. 
In this scenario, there is the possibility of 
realizing a  hierarchical (normal)  or inverted mass spectrum with large mixing.
For large $\Delta$ and   $\Delta'$, a simple hierarchical structure, as suggested by above argument, is unexpected.
To explore more detailed behavior of the
neutrino mass spectrum  and the mixing matrix, 
we need to employ numerical analysis.

To realize reasonable quark and charged lepton mass ratios and the CKM matrix,
it is sufficient to take $\rho_{D}\sim 1$ and $\rho_{U}\sim 1$.
Therefore in the numerical analysis, we first  choose an appropriate setting of weights.
We chose the parameter values  $R = R_{\pm} = 3$, $|r_{H}| = 0.2$, $|r_{H'}| = 0.45$,
$|r_{L}| =|r_{E}| = 0.2$  and considered several values for $r_{T}$ 
and the phases of $r_{L}$, $r_{E}$, $r_{H}$ and $r_{H'}$.
Although the details of numerical analysis are omitted here,
it reveals that the difference between the two cases (i) and (ii) ( 
Eq. (\ref{case1}) and Eq. (\ref{case2})) is not large, except when $\Delta=0,\Delta'=0$.
For $\Delta \geq 2$ and $\Delta'\geq 4$,  
the resulting mass ratios and the MNS matrix exhibit hierarchical behavior like that in the  quark sector.
For small $\Delta$ and $\Delta'$, 
there is no strong dependence on the $r_{T}$ value and the phases of $r_{L}$ and $r_{E}$.
The resulting neutrino mass matrices exhibit the same behavior as expected from the above naive consideration,  
except when $\Delta=0, \Delta'=0$ in case (i).
The two light neutrino masses are close in the realized mass spectrum, 
and the hierarchical (normal) mass spectrum is obtained, except when $\Delta=0, \Delta'=0$ in case (ii).
We investigate the MNS matrix for these values of $\Delta$ and $\Delta'$. 
Variation of the phases of  $r_{H}$ and  $r_{H'}$ modifies the MNS matrix but dose not change the mass ratios.
Typical values of the MNS matrix and the neutrino mass ratios are given as follows: 
\begin{eqnarray}
&&\mbox{For $\Delta = 0$, $\Delta' = 0$ in the case (i), we have} \nonumber \\
&&  \qquad \qquad~~~~~~~
m_{3}/m_{2} \sim  m_{1}/m_{2} \approx {\mathcal O} (10^{-2}) \mbox{~~and} \\ 
&&  \qquad \qquad |U_{\rm MNS}| \sim
\left( \begin{array}{ccc} 
0.9-0.4 &0.4 -0.9 &   0.0  \\  
0.4-0.9 &  0.9 -0.4 &0.4 - 0.6  \\  
 0.0-0.1 & 0.1 & 1.0 
\end{array} \right).\\
&&\mbox{For $\Delta = 2, \Delta'=0$, we have}\nonumber \\
&& \qquad \qquad ~~~~~~~
m_{3}/m_{2} \sim m_{1}/m_{2} \approx {\mathcal O} (10^{-1}) \mbox{~~and}\\ 
&& \qquad \qquad |U_{\rm MNS}| \sim
\left( \begin{array}{ccc} 
0.7 & 0.6 -0.7 &0.3  \\ 
 0.2-0.1 & 0.3-0.2 & 0.9 \\
0.7 & 0.3 -0.1 & 0.1  
\end{array} \right).  
\\
&&\mbox{For $\Delta = 2, \Delta'=2$, we have} \nonumber\\
&& \qquad \qquad ~~~~~~~
 m_{3}/m_{2} \sim m_{1}/m_{2} \approx {\mathcal O} (10^{-1}) \mbox{~~and} \\ 
&&  \qquad\qquad |U_{\rm MNS}| \sim
\left( \begin{array}{ccc} 
0.7-0.6 & 0.5 -0.7 &0.5 - 0.3  \\ 
 0.1-0.3 & 0.5 -0.2 & 0.8 - 0.9 \\
0.7 & 0.6 -0.7 & 0.1-0.2  
\end{array} \right).
\end{eqnarray} 
Here, we have varied the phases  of $r_{H}$ and  $r_{H'}$.
It is obvious that the MNS matrix obtained experimentally \cite{review1,valMNS} 
disfavors the latter two cases, $\Delta=2,\Delta'=0$ and $\Delta=2,\Delta'=2$.
In the result of our numerical analysis, the most feasible case  is $\Delta = 0, \Delta' = 0$, 
where the values of several matrix elements overlap the range of each realistic mixing matrix element. 
This is not completely satisfactory,  because the  value of  $(U_{\rm MNS})_{\tau2}$ is too small.

From the numerical analysis, we find that 
we cannot reproduce a fully realistic mixing matrix based on the simple assumption for the $SU(1,1)$ representation.
However, two points deserve to be stated with regard to  this framework.
First, natural small neutrino masses are induced by the spontaneous generation of generations.
They are obtained by the $SU(2)$ triplets that have huge masses through the horizontal symmetry breaking.
Second, in contrast to the fact that the $SU(1,1)$ symmetry breaking gives a  small mixing to the CKM matrix, 
the MNS matrix can have relatively large mixing.
Considering the present results together, 
it appears that we cannot claim that the simple extension model with the $SU(2)$ triplets based on the $SU(1,1)$ horizontal symmetry explains all the mass ratios for quarks and leptons and both mixing matrices.
It would be more natural to consider a higher-dimensional noncompact group as a horizontal symmetry group or 
some symmetry group  that contains the $SU(1,1)$ horizontal group and the SM gauge group.

\section*{Acknowledgements}
 One of the authors (N. Y.) would like to thank T. Ota for comments on related references and K. Yoshioka for useful discussions.

\end{document}